\documentclass[conference]{IEEEtran}

\IEEEoverridecommandlockouts

\usepackage{cite}
\usepackage{amsmath,amssymb,amsfonts}
\usepackage{algorithmic}
\usepackage{graphicx}
\usepackage{textcomp}
\usepackage{xcolor}
\usepackage{hyperref}
\usepackage{multirow}
\usepackage{balance}
\begin{document}

\title{
Open-Amp: Synthetic Data Framework for Audio Effect Foundation Models
}

\author{\IEEEauthorblockN{
Alec Wright}
\IEEEauthorblockA{\textit{Acoustics and Audio Group} \\
\textit{University of Edinburgh}\\
Edinburgh, UK \\
alec.wright@ed.ac.uk}
\and
\IEEEauthorblockN{
Alistair Carson \thanks{A. Carson is funded by the Scottish Graduate School of Arts and Humanities (SGSAH).}}
\IEEEauthorblockA{\textit{Acoustics and Audio Group} \\
\textit{University of Edinburgh}\\
Edinburgh, UK \\
alistair.carson@ed.ac.uk}
\and
\IEEEauthorblockN{
Lauri Juvela}
\IEEEauthorblockA{\textit{DICE} \\
\textit{Aalto University}\\
Espoo, Finland \\
lauri.juvela@aalto.fi}
}

\maketitle

\begin{abstract}
This paper introduces Open-Amp, a synthetic data framework for generating large-scale and diverse audio effects data. Audio effects are relevant to many musical audio processing and Music Information Retrieval (MIR) tasks, such as modelling of analog audio effects, automatic mixing, tone matching and transcription. Existing audio effects datasets are limited in scope, usually including relatively few audio effects processors and a limited amount of input audio signals. Our proposed framework overcomes these issues, by crowdsourcing neural network emulations of guitar amplifiers and effects, created by users of open-source audio effects emulation software. This allows users of Open-Amp complete control over the input signals to be processed by the effects models, as well as providing high-quality emulations of hundreds of devices. Open-Amp can render audio online during training, allowing great flexibility in data augmentation. Our experiments show that using Open-Amp to train a guitar effects encoder achieves new state-of-the-art results on multiple guitar effects classification tasks. Furthermore, we train a one-to-many guitar effects model using Open-Amp, and  use it to emulate unseen analog effects via manipulation of its learned latent space, indicating transferability to analog guitar effects data.
\end{abstract}


\section{Introduction}
\label{sec:intro}



Digital audio effects modelling is a field of research that aims to emulate analog audio effects processors digitally, allowing musicians to replace analog hardware with software. Popular targets for modelling include phasers \cite{kiiski2016time}, flangers \cite{mavcak2016simulation}, compressors \cite{eichas2016modeling}, distortion pedals \cite{nercessian2021lightweight} and guitar amplifiers \cite{Karjalainen06}. In recent years, machine learning techniques, specifically neural networks, have widely been applied to modelling of guitar amplifiers \cite{damskaggICASSP, WrightRNN19, WrightApplSci, vanhatalo2022review}, and other effects \cite{Martinez2021_Thesis,  peussa2021exposure, wright2021neural, carson2023differentiable, mikkonen2023neural, WrightNeural}. 

One popular application of neural modelling techniques is guitar amplifier profiling or capture. This allows users to create an emulation of a guitar amplifier or pedal, by collecting training data from the target device. There are numerous commercial products available, both software and hardware based, that offer this as a feature, allowing non-expert users to create high-quality emulations for use in music performance and production contexts. 

Open-source software for neural effects modelling is also available, allowing anyone with a personal computer, audio interface and internet connection to capture their own devices. Two popular examples are GuitarML \cite{GuitarML} and Neural Amp Modeler \cite{NAM}. Online communities have formed with users sharing their amp and effects pedal captures. We propose using these captures as part of a synthetic data framework, Open-Amp, augmenting audio that can subsequently be used for downstream tasks. This addresses a gap in the datasets and augmentation frameworks currently available, the lack of large-scale and diverse training data for audio effects. 

Recent work has highlighted the application of audio effects data for a range of tasks, for example in audio effect and mixing style transfer\cite{steinmetz2022style, koo2023music}, audio effect removal \cite{RiceAudioEffectRemoval, TakeAudioChainRemoval, lee2024distortion} and zero-shot guitar amplifier modelling \cite{chen2024towards}. Additionally, it had been shown that for related tasks such as Automatic Music Transcription (AMT) for guitar, adding effects processing such as guitar amplifier emulations to training data can aid in performance and improve generalisation \cite{chen2022towardsegdb, zang2024synthtab}.

Typically audio effects data is either collected from physical devices, either manually \cite{schmitz2018introducing, Hawley2019_SigTrainAES} or using mechanical actuators \cite{juvela2023end}. This however introduces issues related to scaling, as it requires access to and time to gather data from each physical device. More frequently, data is produced by applying digital audio effects as a kind of augmentation. Existing software, however, lacks the diverse range of effects processing devices used by musicians.


The contributions of this paper are as follows:
\begin{itemize}
    \item Release Open-Amp\footnote{\scriptsize\url{https://github.com/Alec-Wright/OpenAmp}} - a Python package which allows for online augmentation of any input audio with a diverse range of guitar amplifier and distortion effects models.
    \item Use Open-Amp to train a guitar effects encoder, and demonstrate transferability of the learned embeddings on downstream classification tasks
    \item Create a one-to-many guitar effects model with Open-Amp, and use its learned latent space to enroll novel analog audio effects processing devices
    
\end{itemize}

With this, we show that Open-Amp is a flexible and efficient way of creating the large-scale and diverse audio effects data that is needed to create audio effect foundation models.





\section{Related Work}
\label{sec:related}

\subsection{Data Augmentation}
\label{ssec:aug}

There are existing packages that apply a range of audio effects, for data augmentation and differentiable DSP purposes. Pedalboard \cite{sobot_peter_2023_7817838} is an audio effects augmentation software package, that offers a range of basic audio effects, including distortion. The Differentiable Audio Signal Processors in Pytorch (DASP) \cite{steinmetz_dasp} and Differentiable Digital Signal Processing (DDSP) \cite{DDSP} libraries offer a range of differentiable audio effects and common audio processors. TorchAudio \cite{hwang2023torchaudio} also offers commonly used audio transformations and effects. 

Whilst these packages have great utility in data augmentation pipelines, they all offer relatively simple implementations of audio effects. In practice, there are a vast range of products available for each class of audio effect that a musician might use. For example, searching any online music gear store for guitar amplifiers will reveal hundreds, if not thousands, of products on offer. These devices all have unique tonal characteristics. In contrast, the existing audio effects augmentation libraries offer relatively limited options for emulating guitar amplifiers, generally providing memoryless nonlinear functions and filters, which could be used to create block-based Wiener-Hammerstein models \cite{eichas:2016}, for example. 

Pedalboard can be used to load audio effects plugins, however this is not always practical for data augmentation purposes for two reasons, firstly, machine learning training is often carried out on Linux servers and the majority of plugins released do not include a Linux build. This does not prevent offline data generation on a different operating system, however it is often desirable to apply data augmentations on-the-fly during training. Secondly, whilst a diverse range of audio effects plugins are available, they are generally commercial products, so not readily available to researchers.

\subsection{Audio Effects Datasets}
\label{ssec:afxdata}

There are several existing datasets intended for use in modelling/classification of audio effects. 
The GUITAR-FX-DIST (GFX) dataset \cite{comunità2021guitar} 
consists of guitar processed with 13 different digital emulations of distortion effects. 
The EGFX dataset 
contains electric guitar processed by 12 different analog audio effects, either distortion, reverb, modulation or delay \cite{pedroza2022egfxset}.
The EGDB dataset 
consists of electric guitar processed by various digital guitar amplifier models, and is intended for the task of automatic transcription \cite{chen2022towardsegdb}. The Amp-Space Dataset \cite{naradowsky2021amp} also contains data collected from a diverse range of devices. However, there is a limited amount of diversity in the input signal processed by the devices.  


Whilst these datasets have been great contributions to the field of machine learning for audio effects, there are a number of challenges and limitations

Firstly, existing datasets choose a collection of input guitar signals, such as single held notes (EGDB, GFX) or guitar playing (EGDB, amp-space), which are subsequently 'baked-in' to the dataset. This limits the domain that the dataset is applicable over. For example, it is possible that guitar effects models trained on single notes, perform poorly on polyphonic and other more complex guitar signals.

Secondly, the diversity of devices is often quite limited, with relatively small collections of devices available, 13 for GFX, 12 for EGFX and 6 for EGDB, and 52 for amp-space. This, as mentioned earlier, is potentially insufficient when compared to the number and range of devices available.

Finally, audio effects generally have multiple parameters. Guitar amplifiers typically have in excess of five parameters that control of timbre of the device. It is desirable to include a sufficient sampling of these parameters in any dataset produced \cite{mikkonen2024sampling}. This is challenging to achieve for analog audio effects, with existing research relying on mechanical automation \cite{naradowsky2021amp, juvela2023end}. Other published datasets either do not vary the effect parameters (EGFX,  EGDB), or are using digital emulations which allow for straightforward control over the effect parameters (GFX).



\section{Experiments}
\label{sec:experiments}

Our proposed data synthesis framework primarily addresses the first two points raised in Sec.~\ref{ssec:afxdata}. The input signal can be any audio signal chosen by the user, as the models themselves are available to render audio. The diversity of devices available is also very great, and likely to increase as more users create and upload models of their devices.

For the experiments presented in this paper, we used the ``Proteus Tone Packs'' collection from the Guitar ML tone library\footnote{\scriptsize\url{https://guitarml.com/tonelibrary/tonelib-pro.html}}. This collection of models consists of 59 guitar amplifier captures and 101 effects pedal captures. Sixty-five of the models include a single parameter, typically the gain, as a conditioning parameter.
It should be noted, however, that additional models can be included using Open-Amp, for example using different architectures, and including even more devices. All models are single-layer LSTMs with hidden size 40 \cite{WrightRNN19, WrightApplSci}.

To demonstrate the transferability of the Open-Amp framework, we use it to train neural network models to achieve two tasks, guitar distortion effects classification, and guitar effects emulation. For all experiments, the data was rendered online during training, utilising multi-process dataloading in PyTorch to ensure minimal batch loading time, with unprocessed electric guitar audio from the IDMT-Guitar transcription dataset \cite{kehling2014automatic} as input. 

\subsection{Guitar Effects Classification}

We trained a guitar effects encoder on data synthesised by Open-Amp, using a contrastive framework based on SimCLR \cite{chen2020simple}. The encoder architecture is similar to the 'Music Effects Encoder' used in \cite{koo2022end}. It consists of one-dimensional convolutional blocks, with each block containing two convolutional layers with residual connection. Each convolutional layer is followed by a batch normalization layer and a ReLU activation. The output convolutional layer has 64 channels, and the encoding is averaged over time to produce the final embedding with 64 dimensions. The encoder has six blocks with kernel size of five and channels growing progressively from 16 to 64, giving a total of 112888 learnable parameters.

To generate each training batch, random 1-second clips are sampled from the input data. For each device, a positive pair consists of two different clips from the input dataset, processed by one of the effects models. This is repeated for a random selection of the available models. Negative examples then consists of all examples in the batch that were processed by different effects models. We used a batch size of 128 and trained for 200,000 iterations using the normalized temperature-scaled cross-entropy loss \cite{chen2020simple}. 

We evaluate our contrastively trained encoder on an existing classification task from the GFX dataset, using results from the original paper as a baseline \cite{comunità2021guitar}. This dataset consists of thirteen different distortion effects, applied to clean guitar. The effects are either applied to mono guitar (single string) or polyphonic guitar (two strings), and the effect parameter settings are either sampled from a discrete distribution of four possible values, or continuously, creating four subsets, Mono Continuous, Mono Discrete, Poly Continuous and Poly Discrete. In the original paper introducing this dataset, a convolutional neural network, FxNet, was trained to classify the effect class, with various combinations of training and test dataset presented to test generalisation. 

We use our contrastively trained encoder to extract embeddings from the GFX dataset. We create two classifiers based on these embeddings, a k-nearest neighbours (KNN) classifier and a single layer MLP with hidden size 100 and Rectified Linear Unit (ReLU) non-linearity. We report the classifier accuracy in Table~\ref{tab:acc_gfx}, and compare it to the accuracy reported for the FxNet presented in \cite{comunità2021guitar}. Note that our contrastively trained encoder has not seen any data from the GFX dataset during training. For the KNN classifier, the performance is worse than the baseline, however typically by a small margin. The single-layer MLP however, is able to achieve better accuracy than the baseline FxNet in six out of eight cases. For comparison, the FxNet has a reported 760,000 learnable parameters, compared to our encoder which has just 112,888. This demonstrates the transferability of the encoder's learned embeddings, and thus of Open-Amp, to downstream tasks.

\begin{table}[htb]
\caption{Classifier accuracy on GFX \cite{comunità2021guitar} dataset.}
\label{tab:acc_gfx}
\begin{tabular}{cc|ccc}

\multicolumn{2}{c|}{Classifier Accuracy}                      & \multicolumn{3}{c}{Model}                                                                                                                                                                                                                      \\ \hline
                                                 &            & \multirow{2}{*}{\begin{tabular}[c]{@{}c@{}}FxNet \\ (Baseline)\end{tabular}} & \multirow{2}{*}{\begin{tabular}[c]{@{}c@{}}Open-Amp\\ KNN (ours)\end{tabular}} & \multirow{2}{*}{\begin{tabular}[c]{@{}c@{}}Open-Amp\\ MLP (ours)\end{tabular}} \\
Train Set                                        & Test Set   &                                                                              &                                                                                &                                                                                \\ \hline
\multicolumn{1}{c|}{Mono Disc.}                  & Mono Disc. & 86.3                                                                         & 83.0                                                                           & \textbf{89.3}                                                                  \\
\multicolumn{1}{c|}{}                            & Mono Cont. & 83.1                                                                         & 81.0                                                                           & \textbf{86.7}                                                                  \\ \hline
\multicolumn{1}{c|}{\multirow{2}{*}{Mono Cont.}} & Mono Disc. & \textbf{81.3}                                                                & 76.3                                                                           & 81.2                                                                           \\
\multicolumn{1}{c|}{}                            & Mono Cont. & 90.9                                                                         & 89.3                                                                           & \textbf{91.9}                                                                  \\ \hline
\multicolumn{1}{c|}{\multirow{2}{*}{Poly Disc.}} & Poly Disc. & 88.4                                                                         & 86.7                                                                           & \textbf{89.4}                                                                  \\
\multicolumn{1}{c|}{}                            & Poly Cont. & \textbf{89.4}                                                                & 85.3                                                                           & 86.5                                                                           \\ \hline
\multicolumn{1}{c|}{\multirow{2}{*}{Poly Cont.}} & Poly Disc. & 84.1                                                                         & 82.7                                                                           & \textbf{85.4}                                                                  \\
\multicolumn{1}{c|}{}                            & Poly Cont. & 91.4                                                                         & 91.4                                                                           & \textbf{93.0}                                                                  \\ \hline
\multicolumn{2}{c|}{Overall}                                  & 86.9                                                                         & 84.5                                                                           & \textbf{87.9}                                                                 
\end{tabular}
\end{table}

As a further ablation we carry out additional classification tasks, on four different datasets, the previously introduced EGFX, EGDB and GFX datasets (see Sec.~\ref{ssec:afxdata}), as well as ``Open30'', a dataset consisting of guitar audio processed by 30 randomly selected devices from the Proteus Tone Pack, synthesised using Open-Amp. This dataset contains 1104 two-second clips from each target device, using data from the IDMT transcription dataset \cite{kehling2014automatic} as input. We then train three additional encoders using the contrastive framework described earlier on the previously introduced datasets, EGFX, EGDB and GFX. These encoders were trained for 100,000 iterations, with a batch size equal to the number of guitar effects contained in each dataset. These three encoders, in addition to a version of the Open-Amp encoder introduced earlier that was only trained for 100,000 iterations, were then used to extract encodings over the four datasets. These encodings were then used to fit a single-layer MLP classifier, with various permutations of training and test set. If the training and test sets are from the same dataset, a random class-balanced train-test split of 85-15 was used. The results are shown in Table \ref{tab:acc_ablation_mlp}. The results show that the encoder trained using Open-Amp performs best on all datasets, further demonstrating transferability to unseen data.



\begin{table}[htb]
\caption{Cross-dataset evaluation of guitar effects encoders.}
\label{tab:acc_ablation_mlp}
\centering
\begin{tabular}{cc|cccc}
\multicolumn{2}{c|}{\multirow{2}{*}{\begin{tabular}[c]{@{}c@{}}Effect Class Accuracy (\%)\\  - MLP Classifier\end{tabular}}} & \multicolumn{4}{c}{Test Set}                                  \\ \cline{3-6} 
\multicolumn{2}{c|}{}                                                                                                        & GFX           & EGFX          & EGDB          & Open30      \\ \hline
\multicolumn{1}{c|}{\multirow{4}{*}{\begin{tabular}[c]{@{}c@{}}Train\\ Set\end{tabular}}}             & GFX                  & 91.2          & 68.3          & 72.3          & 96.0          \\
\multicolumn{1}{c|}{}                                                                                 & EGFX                 & 57.4          & 64.4          & 45.9          & 51.0          \\
\multicolumn{1}{c|}{}                                                                                 & EGDB                 & 89.7          & 69.2          & 75.7          & 92.2          \\
\multicolumn{1}{c|}{}                                                                                 & Open-Amp             & \textbf{91.8} & \textbf{72.0} & \textbf{91.1} & \textbf{99.8}
\end{tabular}
\vspace{-4mm}
\end{table}

\subsection{Guitar Effect Emulation}
Here we explore the application of Open-Amp to a one-to-many guitar effect emulation task. Recent work investigated this problem using a propriety dataset \cite{chen2024towards}, but we demonstrate that a similar model can be trained using Open-Amp. This involved two stages: training a \textit{foundation model} on Open-Amp synthetic devices then \textit{enrolling} unseen analog devices into the model. Open-Amp effects models with a conditioning parameter were treated as five separate models, with the conditioning value linearly spaced from 0 to 1, resulting in a total of 394 synthetic devices.

\subsubsection{Foundation model}
\begin{figure}[t!]
    \centering
    \includegraphics[width=0.5\textwidth, trim={0, 1.0cm, 0, 1.0cm}]{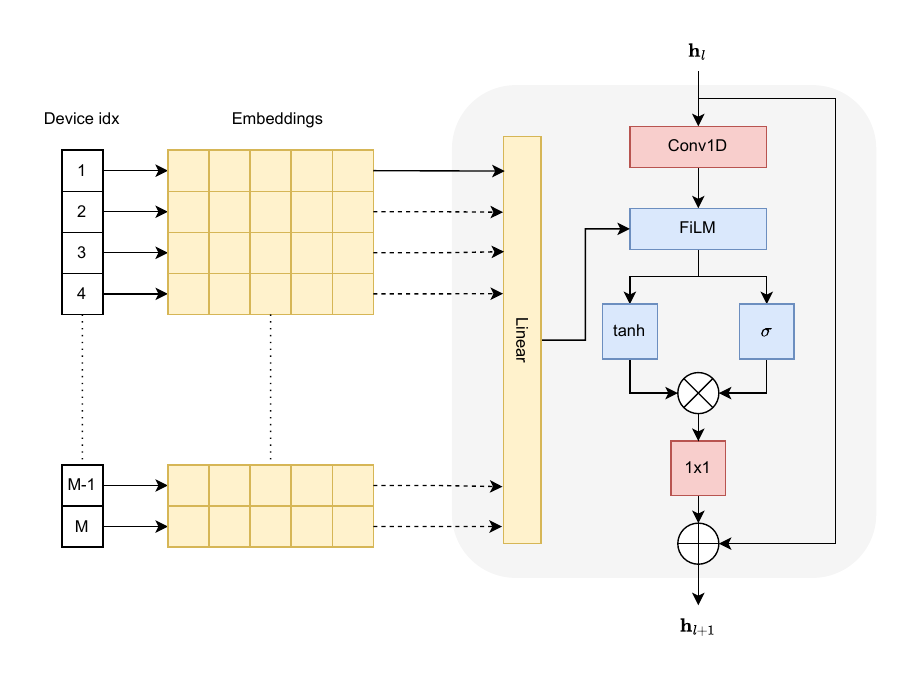}
    \caption{TCN layer for the one-to-many guitar effect emulation task. ${\bf h}_l \in \mathbb{Re}^{N \times C \times T}$ is the output from the $l$th layer where $N$ is the batch size, $C$ is the channel width and $T$ is the time-dimension. $M$ is the number of devices used during training and the embedding dimension is $E=5$ here for illustrative purposes. This example shows a forward pass conditioned on the 1\textsuperscript{st} device embedding. The learnable embeddings are shared amongst all layers. }
    \label{fig:tcn}
        \vspace{-4mm}
\end{figure}

    The foundation model architecture was a temporal convolutional network (TCN) conditioned with feature-wise linear modulation (FiLM) \cite{chen2024towards, steinmetz2022_tcn, Comunita2022}. A single layer is shown in Fig. \ref{fig:tcn}. This consisted of 2 TCN blocks each with $L=8$ layers with $C=16$ convolutional channels, a kernel size of $K=3$ and dilation growth $D=2$. The model was conditioned on a learnable look-up table with a unique embedding for each of the $M=394$ (synthetic) devices seen during training. We trained three versions of the model with embedding dimensions of $E = \{16, 64, 256\}$. The model was trained on an input signal of 35 minutes of (clean) audio data, giving a total of 230 hours of synthetically generated training data. The data was segmented into 2s clips, with batch size of $N=16$, and the model trained for 8 epochs (approximately 60 hours on an NVIDIA GeForce GTX 1080). The loss function was the sum of the error-to-signal ratio (ESR) \cite{damskaggICASSP} and multi-resolution spectral loss (MRSL) \cite{steinmetz2020auraloss}.

    The results on unseen test audio can be seen in Table~ \ref{tab:fx_foundation_db}. Five devices from the training set were selected to show the spread of results: the best, 25th percentile, median, 75th percentile and worst performing devices from the Emb-64 model (in terms of combined ESR and MRSL). A baseline one-to-one TCN was trained for each device, with the results shown in the first column. For all devices the one-to-one model outperforms the one-to-many models, as was observed in \cite{chen2024towards}. In many cases however, the best performing one-to-many model (Emb-256) provides comparable results to the baseline and is at most 5dB (ESR) worse for the devices presented. Generally the results improve as the embedding dimension increases.

\begingroup
\vspace{-3mm}
\setlength{\tabcolsep}{4pt} 
\begin{table}[htb]
\caption{Foundation model test loss (in decibels) for selected devices, bold indicates best performing non-baseline model}
\label{tab:fx_foundation_db}
\centering
{
\begin{tabular}{c|cc|cc|cc|cc}
  & \multicolumn{2}{|c|}{1-to-1 (base)}& \multicolumn{2}{|c|}{Emb-16} & \multicolumn{2}{|c|}{Emb-64}  & \multicolumn{2}{|c}{Emb-256}  \\
 \hline
 Dev. & ESR & MRSL & ESR & MRSL & ESR & MRSL & ESR & MRSL \\
\hline
1 &-24.1&-8.0& -19.2&-5.9&-23.2&-7.0&{\bf -23.7}&{\bf -7.4} \\
2 &-29.4&-6.0& -18.6&-3.3&-18.6&-3.6&{\bf-24.0}&{\bf -4.6} \\
3 &-20.6&-3.9& -12.4&-1.9&-14.8&{\bf -2.7}&{\bf -15.8}&-2.5 \\
4&-16.5&-2.5& -11.1&-0.9&-12.8&-1.5&{\bf -13.1}&{\bf -1.7} \\
5&-6.7&-0.9& -0.7&1.1&{\bf -2.4}&{\bf 0.5}&-2.0&0.7 \\
\end{tabular}
}
\vspace{-2.5mm}
\end{table}
\endgroup

\subsubsection{Unseen device enrolment}
To test the generality of the learned embedding space, we introduce unseen analog audio effects to the foundation models: namely three distortion/overdrive pedals from the EGFX database \cite{pedroza2022egfxset}. The database contains 57 minutes of clean/processed audio for each device, from which we select a 90-5-5 train-validation-test split. For each device, we freeze all parameters in the foundation model except the embedding space, and fine-tune the model in a one-to-one fashion, therefore teaching the model to learn a new embedding for the unseen device. The initial embedding was set to the Open-Amp embedding which gave the lowest loss over the unseen data. As a baseline we train a fully-learnable one-to-one TCN for each device. We run the experiments on different subsets of the full training dataset, ranging from 3s (0.1\%) to 53 minutes (100\%). 

\begin{figure}[]
    \centering
    \includegraphics[width=0.5\textwidth, trim={0, 0.5cm, 0, 0cm}]{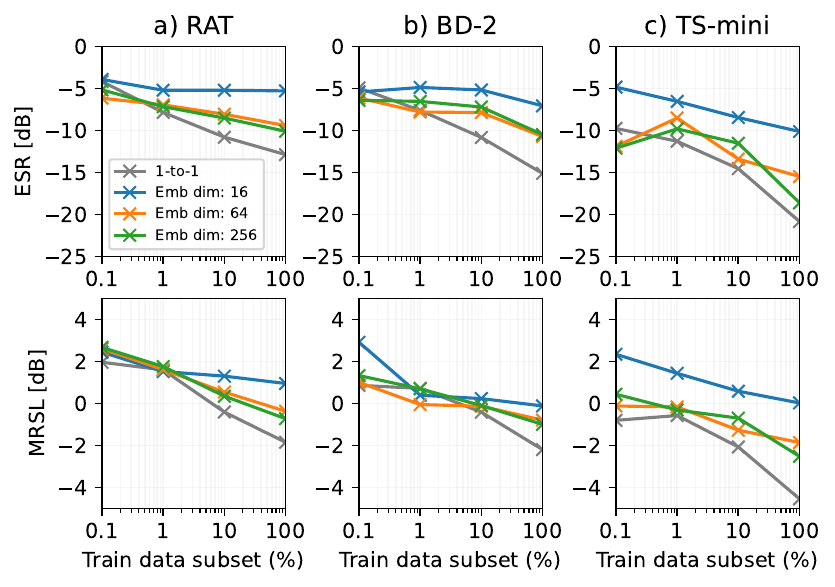}
    \caption{Enrolment experiment results for unseen devices from the EGFX dataset: a) Proco Rat b) Boss Blues Driver and c) Tube Screamer Mini. The grey line shows the test loss of a device-specific one-to-one TCN model. }
    \label{fig:enrolment}
    \vspace{-4mm}
\end{figure}

The results in Fig. \ref{fig:enrolment} show that the proposed embedding enrolment method can give comparable results to training a one-to-one model, especially for higher embedding dimensions and fewer training data. In many cases the discrepancies between the baseline and the embedding models are similar to those seen in Table \ref{tab:fx_foundation_db}, indicating that the embedding space learned by the model from synthetic data can represent unseen analog audio effects to a similar degree of accuracy. As the duration of training data decreases, the performance of the proposed models converge with that of the baseline and in some case does slightly better. This suggests that this foundation-enrolment approach may be useful when training data scarcity is an issue.


\section{Conclusion}
\label{sec:conclusion}

In this work, we presented Open-Amp, a large-scale synthetic audio effects augmentation framework. Utilising crowdsourced models from open-source neural effect modelling software, Open-Amp allows for flexible online data augmentation with a diverse range of realistic distortion effect models. Our guitar effects encoder, trained with Open-Amp, outperforms classifiers trained on domain-specific audio effect datasets across various audio effects classification tasks, indicating that Open-Amp augmentation generalises well to unseen effects. 
Furthermore, we use Open-Amp to train a one-to-many guitar effects model, and use the learned latent space for enrolment of unseen devices. Our framework is suitable for online effects augmentation, enabling its integration into data augmentation pipelines and making it valuable for creating foundation models in audio effects that can be used for a wide range of machine learning tasks. Future work will involve expanding Open-Amp to include more effects, adding variable sample rates to the models \cite{carson2024sample}, as well as adding incorporating white-box models of common circuits, such as the tone-stack \cite{yeh2006discretization, miklanek2023neural}, to further expand the range of augmentations available.


\clearpage
\balance
\bibliographystyle{IEEEtran}
\bibliography{refs}

\begin{thebibliography}{10}
\providecommand{\url}[1]{#1}
\csname url@samestyle\endcsname
\providecommand{\newblock}{\relax}
\providecommand{\bibinfo}[2]{#2}
\providecommand{\BIBentrySTDinterwordspacing}{\spaceskip=0pt\relax}
\providecommand{\BIBentryALTinterwordstretchfactor}{4}
\providecommand{\BIBentryALTinterwordspacing}{\spaceskip=\fontdimen2\font plus
\BIBentryALTinterwordstretchfactor\fontdimen3\font minus
  \fontdimen4\font\relax}
\providecommand{\BIBforeignlanguage}[2]{{%
\expandafter\ifx\csname l@#1\endcsname\relax
\typeout{** WARNING: IEEEtran.bst: No hyphenation pattern has been}%
\typeout{** loaded for the language `#1'. Using the pattern for}%
\typeout{** the default language instead.}%
\else
\language=\csname l@#1\endcsname
\fi
#2}}
\providecommand{\BIBdecl}{\relax}
\BIBdecl

\bibitem{kiiski2016time}
R.~Kiiski, F.~Esqueda, and V.~V{\"a}lim{\"a}ki, ``Time-variant gray-box
  modeling of a phaser pedal,'' in \emph{Proc. Int. Conf. Digital Audio Effects
  (DAFx)}, Brno, Czech Republic, Sep 2016, pp. 31--38.

\bibitem{mavcak2016simulation}
J.~Ma{\v{c}}{\'a}k, ``Simulation of analog flanger effect using bbd circuit,''
  in \emph{Proc. Int. Conf. Digital Audio Effects (DAFx)}, Brno, Czech
  Republic, Sep 2016, pp. 99--102.

\bibitem{eichas2016modeling}
F.~Eichas and U.~Z{\"o}lzer, ``Modeling of an optocoupler-based audio dynamic
  range control circuit,'' in \emph{Novel Optical Systems Design and
  Optimization XIX}, vol. 9948.\hskip 1em plus 0.5em minus 0.4em\relax SPIE,
  2016, pp. 47--62.

\bibitem{nercessian2021lightweight}
S.~Nercessian, A.~Sarroff, and K.~J. Werner, ``Lightweight and interpretable
  neural modeling of an audio distortion effect using hyperconditioned
  differentiable biquads,'' in \emph{Proc. IEEE ICASSP}, Toronto, Canada, June
  2021, pp. 890--894.

\bibitem{Karjalainen06}
M.~Karjalainen and J.~Pakarinen, ``Wave digital simulation of a vacuum-tube
  amplifier,'' in \emph{Proc. IEEE ICASSP}, Toulouse, France, May 2006, pp.
  153--156.

\bibitem{damskaggICASSP}
E.-P. Damsk{\"a}gg, L.~Juvela, E.~Thuillier, and V.~V{\"a}lim{\"a}ki, ``Deep
  learning for tube amplifier emulation,'' in \emph{Proc. IEEE ICASSP},
  Brighton, UK, May 2019, pp. 471--475.

\bibitem{WrightRNN19}
A.~Wright, E.-P. Damsk{\"{a}}gg, and V.~V{\"{a}}lim{\"{a}}ki, ``Real-time
  black-box modelling with recurrent neural networks,'' in \emph{Proc. Int.
  Conf. Digital Audio Effects (DAFx)}, Birmingham, UK, Sep. 2019, pp. 173--180.

\bibitem{WrightApplSci}
A.~Wright, E.-P. Damsk{\"{a}}gg, L.~Juvela, and V.~V{\"{a}}lim{\"{a}}ki,
  ``Real-time guitar amplifier emulation with deep learning,'' \emph{Appl.
  Sci.}, vol.~10, no.~3, 2020.

\bibitem{vanhatalo2022review}
T.~Vanhatalo, P.~Legrand, M.~Desainte-Catherine \emph{et~al.}, ``A review of
  neural network-based emulation of guitar amplifiers,'' \emph{Appl. Sci.},
  vol.~12, no.~12, p. 5894, 2022.

\bibitem{Martinez2021_Thesis}
M.~A. Martínez~Ramírez, ``Deep learning for audio effects modeling,'' Ph.D.
  dissertation, Queen Mary University of London, London, UK, Nov. 2020.

\bibitem{peussa2021exposure}
A.~Peussa, E.-P. Damsk{\"a}gg, T.~Sherson, S.~I. Mimilakis, L.~Juvela,
  A.~Gotsopoulos, and V.~V{\"a}lim{\"a}ki, ``Exposure bias and state matching
  in recurrent neural network virtual analog models,'' in \emph{Proc. Int.
  Conf. Digital Audio Effects (DAFx)}, 2021, pp. 284--291.

\bibitem{wright2021neural}
A.~Wright and V.~V{\"a}lim{\"a}ki, ``Neural modeling of phaser and flanging
  effects,'' \emph{J. Audio Eng. Soc.}, vol.~69, no. 7/8, pp. 517--529, Jul.
  2021.

\bibitem{carson2023differentiable}
A.~Carson, C.~Valentini-Botinhao, S.~King, and S.~Bilbao, ``Differentiable
  grey-box modelling of phaser effects using frame-based spectral processing,''
  in \emph{Proc. Int. Conf. Digital Audio Effects (DAFx)}, Copenhagen, Denmark,
  Sep 2023, pp. 219--226.

\bibitem{mikkonen2023neural}
O.~Mikkonen, A.~Wright, E.~Moliner, and V.~V{\"a}lim{\"a}ki, ``Neural modeling
  of magnetic tape recorders,'' in \emph{Proc. Int. Conf. Digital Audio Effects
  (DAFx)}, Copenhagen, Denmark, Sep 2023, pp. 196--203.

\bibitem{WrightNeural}
A.~Wright, ``Neural modelling of audio effects,'' Ph.D. dissertation, Aalto
  University, Espoo, Finland, Dec 2023.

\bibitem{GuitarML}
\BIBentryALTinterwordspacing
K.~Bloemer, ``{GuitarML}.'' [Online]. Available: \url{https://guitarml.com/}
\BIBentrySTDinterwordspacing

\bibitem{NAM}
\BIBentryALTinterwordspacing
S.~Atkinson, ``Neural amp modeler.'' [Online]. Available:
  \url{https://www.neuralampmodeler.com/}
\BIBentrySTDinterwordspacing

\bibitem{steinmetz2022style}
C.~J. Steinmetz, N.~J. Bryan, and J.~D. Reiss, ``Style transfer of audio
  effects with differentiable signal processing,'' \emph{J. Audio Eng. Soc.},
  vol.~70, no.~9, pp. 708--721, Sep. 2022.

\bibitem{koo2023music}
J.~Koo, M.~A. Mart{\'\i}nez-Ram{\'\i}rez, W.-H. Liao, S.~Uhlich, K.~Lee, and
  Y.~Mitsufuji, ``Music mixing style transfer: A contrastive learning approach
  to disentangle audio effects,'' in \emph{Proc. IEEE ICASSP}, 2023, pp. 1--5.

\bibitem{RiceAudioEffectRemoval}
M.~Rice, C.~J. Steinmetz, G.~Fazekas, and J.~D. Reiss, ``General purpose audio
  effect removal,'' in \emph{Proc. IEEE Workshop on Applications of Signal
  Processing to Audio and Acoustics (WASPAA)}, 2023, pp. 1--5.

\bibitem{TakeAudioChainRemoval}
O.~Take, K.~Watanabe, T.~Nakatsuka, T.~Cheng, T.~Nakano, M.~Goto, S.~Takamichi,
  and H.~Saruwatari, ``Audio effect chain estimation and dry signal recovery
  from multi-effect-processed musical signals,'' in \emph{Proc. Int. Conf.
  Digital Audio Effects (DAFx)}, Guildford, Surrey, Sep 2024, pp. 1--8.

\bibitem{lee2024distortion}
Y.-S. Lee, Y.-P. Peng, J.-T. Wu, M.~Cheng, L.~Su, and Y.-H. Yang, ``Distortion
  recovery: A two-stage method for guitar effect removal,'' in \emph{Proc. Int.
  Conf. Digital Audio Effects (DAFx)}, Guildford, United Kingdom, Sep 2024, pp.
  177--184.

\bibitem{chen2024towards}
Y.-H. Chen, Y.-T. Yeh, Y.-C. Cheng, J.-T. Wu, Y.-H. Ho, J.-S.~R. Jang, and
  Y.-H. Yang, ``Towards zero-shot amplifier modeling: One-to-many amplifier
  modeling via tone embedding control,'' \emph{arXiv preprint
  arXiv:2407.10646}, 2024.

\bibitem{chen2022towardsegdb}
Y.-H. Chen, W.-Y. Hsiao, T.-K. Hsieh, J.-S.~R. Jang, and Y.-H. Yang, ``Towards
  automatic transcription of polyphonic electric guitar music: A new dataset
  and a multi-loss transformer model,'' in \emph{Proc. IEEE ICASSP}, 2022, pp.
  786--790.

\bibitem{zang2024synthtab}
Y.~Zang, Y.~Zhong, F.~Cwitkowitz, and Z.~Duan, ``Synthtab: Leveraging
  synthesized data for guitar tablature transcription,'' in \emph{Proc. IEEE
  ICASSP}, 2024, pp. 1286--1290.

\bibitem{schmitz2018introducing}
T.~Schmitz and J.-J. Embrechts, ``Introducing a dataset of guitar amplifier
  sounds for nonlinear emulation benchmarking,'' in \emph{Proc. 145th Audio
  Eng. Soc. Conv.}\hskip 1em plus 0.5em minus 0.4em\relax New York, USA, Oct.
  2018.

\bibitem{Hawley2019_SigTrainAES}
S.~Hawley, B.~Colburn, and S.~I. Mimilakis, ``Profiling audio compressors with
  deep neural networks,'' in \emph{Proc. 147th Audio Eng, Soc. Conv.}, New
  York, USA, Oct. 2019, Conference Paper.

\bibitem{juvela2023end}
L.~Juvela, E.-P. Damsk{\"a}gg, A.~Peussa, J.~M{\"a}kinen, T.~Sherson, S.~I.
  Mimilakis, K.~Rauhanen, and A.~Gotsopoulos, ``End-to-end amp modeling: from
  data to controllable guitar amplifier models,'' in \emph{Proc. IEEE ICASSP},
  2023, pp. 1--5.

\bibitem{sobot_peter_2023_7817838}
\BIBentryALTinterwordspacing
P.~Sobot, ``Pedalboard,'' Jul. 2021. [Online]. Available:
  \url{https://doi.org/10.5281/zenodo.7817838}
\BIBentrySTDinterwordspacing

\bibitem{steinmetz_dasp}
\BIBentryALTinterwordspacing
C.~Steinmetz, ``Differentiable audio signal processors in pytorch,'' Jan. 2023.
  [Online]. Available: \url{https://github.com/csteinmetz1/dasp-pytorch}
\BIBentrySTDinterwordspacing

\bibitem{DDSP}
J.~Engel, L.~Hantrakul, C.~Gu, and A.~Roberts, ``{DDSP:} {D}ifferentiable
  digital signal processing,'' in \emph{Proc. Int. Conf. Learning
  Representations (ICLR)}, Addis Ababa, Ethiopia, 2020.

\bibitem{hwang2023torchaudio}
J.~Hwang, M.~Hira, C.~Chen, X.~Zhang, Z.~Ni, G.~Sun, P.~Ma, R.~Huang,
  V.~Pratap, Y.~Zhang, A.~Kumar, C.-Y. Yu, C.~Zhu, C.~Liu, J.~Kahn,
  M.~Ravanelli, P.~Sun, S.~Watanabe, Y.~Shi, Y.~Tao, R.~Scheibler, S.~Cornell,
  S.~Kim, and S.~Petridis, ``Torchaudio 2.1: Advancing speech recognition,
  self-supervised learning, and audio processing components for pytorch,''
  2023.

\bibitem{eichas:2016}
F.~Eichas and U.~Z{\"o}lzer, ``Black-box modeling of distortion circuits with
  block-oriented models,'' in \emph{Proc. Int. Conf. Digital Audio Effects
  (DAFx-16)}, Brno, Czech Republic, Sept. 2016, pp. 39--45.

\bibitem{comunità2021guitar}
M.~Comunità, D.~Stowell, and J.~D. Reiss, ``Guitar effects recognition and
  parameter estimation with convolutional neural networks,'' \emph{J. Audio
  Eng. Soc.}, vol.~69, no. 7/8, pp. 594--604, July 2021.

\bibitem{pedroza2022egfxset}
H.~Pedroza, G.~Meza, and I.~R. Roman, ``{EGFxSet}: Electric guitar tones
  processed through real effects of distortion, modulation, delay and reverb,''
  \emph{ISMIR Late Breaking Demo}, 2022.

\bibitem{naradowsky2021amp}
J.~Naradowsky, ``Amp-space: A large-scale dataset for fine-grained timbre
  transformation,'' in \emph{Proc. Int. Conf. Digital Audio Effects (DAFx)},
  Vienna, Austria, Sep. 2021, pp. 57--64.

\bibitem{mikkonen2024sampling}
O.~Mikkonen, A.~Wright, and V.~V{\"a}lim{\"a}ki, ``Sampling the user controls
  in neural modeling of audio devices,'' \emph{EURASIP J. Audio, Speech, and
  Music Processing}, vol. 2024, no.~1, p.~26, 2024.

\bibitem{kehling2014automatic}
C.~Kehling, J.~Abe{\ss}er, C.~Dittmar, and G.~Schuller, ``Automatic tablature
  transcription of electric guitar recordings by estimation of score- and
  instrument-related parameters,'' in \emph{Proc. Int. Conf. Digital Audio
  Effects (DAFx)}, Erlangen, Germany, Sep. 2014, pp. 219--226.

\bibitem{chen2020simple}
T.~Chen, S.~Kornblith, M.~Norouzi, and G.~Hinton, ``A simple framework for
  contrastive learning of visual representations,'' in \emph{Proc. Int. Conf.
  Machine Learning}.\hskip 1em plus 0.5em minus 0.4em\relax PMLR, 2020, pp.
  1597--1607.

\bibitem{koo2022end}
J.~Koo, S.~Paik, and K.~Lee, ``End-to-end music remastering system using
  self-supervised and adversarial training,'' in \emph{Proc. IEEE ICASSP},
  2022, pp. 4608--4612.

\bibitem{steinmetz2022_tcn}
C.~Steinmetz and J.~Reiss, ``Efficient neural networks for real-time modeling
  of analog dynamic range compression,'' in \emph{Proc. 152nd Audio Eng. Soc.
  Conv.}, The Hague, Netherlands, May 2022.

\bibitem{Comunita2022}
M.~Comunità, C.~Steinmetz, H.~Phan, and J.~Reiss, ``Modelling black-box audio
  effects with time-varying feature modulation,'' in \emph{Proc. IEEE ICASSP},
  Rhodes, Greece, 2023.

\bibitem{steinmetz2020auraloss}
C.~J. Steinmetz and J.~D. Reiss, ``auraloss: {A}udio focused loss functions in
  {PyTorch},'' in \emph{Digital Music Research Network One-day Workshop
  (DMRN+15)}, 2020.

\bibitem{carson2024sample}
A.~Carson, A.~Wright, V.~V{\"a}lim{\"a}ki, J.~Chowdhury, and S.~Bilbao,
  ``Sample rate independent recurrent neural networks for audio effects
  processing,'' in \emph{Proc. Int. Conf. Digital Audio Effects (DAFx)},
  Guildford, Surrey, Sep 2024, pp. 17--24.

\bibitem{yeh2006discretization}
D.~T. Yeh and J.~O. Smith, ``Discretization of the’59 fender bassman tone
  stack,'' in \emph{Proc. Int. Conf. Digital Audio Effects (DAFx)}, Montreal,
  Canada, 2006, pp. 1--8.

\bibitem{miklanek2023neural}
{\v{S}}.~Mikl\'{a}nek, A.~Wright, V.~V{\"a}lim{\"a}ki, and J.~Schimmel,
  ``Neural grey-box guitar amplifier modelling with limited data,'' in
  \emph{Proc. Int. Conf. Digital Audio Effects (DAFx)}, Copenhagen, Denmark,
  Sep 2023, pp. 151--158.

\end{thebibliography}


\end{document}